\documentclass{PoS}
\newcommand\cv{c_{\scriptscriptstyle V}}

\def\ie{{\sl i.e.\/}}
\newcommand\bef{\begin{figure}}
\newcommand\eef[1]{\label{fg:#1}\end{figure}}
\newcommand\beq{\begin{equation}}
\newcommand\eeq[1]{\label{#1}\end{equation}}
\newcommand\beqa{\begin{eqnarray}}
\newcommand\eeqa[1]{\label{#1}\end{eqnarray}}
\newcommand\bet{\begin{table}}
\newcommand\eet[1]{\label{tb:#1}\end{table}}

\title{Lambda Phenomena:\\The Lambda points of liquid Helium and chiral QCD}

\ShortTitle{Lambda phenomena}

\author{\speaker{Sourendu Gupta}\\
        TIFR Mumbai\\
        E-mail: \email{sgupta@tifr.res.in}}

\author{Rishi Sharma\\
        TIFR Mumbai\\
        E-mail: \email{rishi@theory.tifr.res.in}}

\abstract{
 The superfluid transition of liquid Helium shares an interesting
 phenomenon with the chiral limit of QCD: the specific heat, $\cv$,
 is finite at the critical point, but has a cusp. From this follows
 an interesting mixture of universal and non-universal features at
 the critical point. Through the CP symmetry of chiral QCD, this has
 implications for the fourth order baryon number susceptibility,
 $\chi_B^4$, and susceptibilities of higher orders.  Investigations
 of such a scaling will show us whether O(4) scaling is an accurate
 description of baryon-free QCD when the pion mass is realistic.
}

\FullConference{9th International Workshop on Critical Point and Onset of Deconfinement - CPOD2014,\\
		17-21 November 2014\\
		ZiF (Center of Interdisciplinary Research), University of Bielefeld, Germany}

\begin{document}

\section{Introduction}

In the chiral limit of QCD, obtained by setting the light quark masses to
zero, the global symmetry of QCD is chiral $SU(2)\times SU(2)$. This is
homomorphic to a O(4) symmetry. The universality conjecture then leads us
expect that the critical indices and critical amplitude ratios in chiral
QCD should be the same as for O(4) magnetic systems \cite{pisarski}.

In recent times, the chiral condensate of QCD (which is equivalent to
the spontaneous magnetization of the magnet), its derivatives with
respect to the quark mass (equivalent to the magnetic susceptibility),
and their scaling towards the chiral limit, have been studied with
somewhat ambiguous results \cite{karsch}.

The scaling of the internal energy and the specific heat of O(N)
symmetric systems is intricate, as is known from the phenomenology of
liquid Helium. However, with the extensive lattice QCD computations now
available on quark number susceptibilities in QCD at zero baryon density
\cite{qns}, their important role in heavy-ion collisions \cite{ilgti},
and their connection with temperature derivatives of the free energy, it
is important to initiate the scaling analysis of these quantities. That
is the purpose of this talk.

\section{Scaling and the limits of universality}

The thermodynamics of QCD is characterized by a free energy, which
is a function of some number of intensive control parameters. These
could include the temperature, $T$, and the baryon chemical potential,
$\mu$. If the pion mass were exactly vanishing, then QCD would have
a O(4) global chiral symmetry.  Since we are interested in real QCD,
where the pion is not massless, an explicit chiral symmetry breaking
parameter is needed.  This is the quark mass, $m$, which plays the same
role in QCD as a magnetic field does for the O(4) magnet.

Near the critical point one can decompose the free energy, $F(T,m)$, into
the sum of two terms. One of these is a regular part, $F_r(T,m)$, and the
other is a singular part, $F_s(T,m)$. $F_r(T,m)$ is Taylor expandable around
the critical point, $T=T_c$ and $m=0$ with some large radius of convergence.
The modern theory of critical phenomena starts from the observation that the
most singular part is a scaling function
\beq
   F_s(T,m) = t^{2-\alpha}\Phi(\tau), \quad{\rm where}\quad
     t = \left|1-\frac T{T_c}\right|,\;
     \tau = \frac t{(m/M_0)^{1/\Delta}},
\eeq{scaling}
where we have chosen the scaling variables $t$ and $\tau$ to be
dimensionless, $T_c$ is the critical temperature, $M_0$ is any
mass scale which remains finite in the chiral limit, and $\alpha$,
$\Delta=\beta\delta$ are critical exponents.  The function $F_s(t,\tau)$,
defined so, is universal, in the sense that whether we examine an O(4)
Heisenberg magnet, QCD, a non-linear sigma model of pions, or the
Nambu-Jona-Lasinio (NJL) model, the $F_s(t,\tau)$ we obtain from all
of them are the same. As a result, the universal properties of thermal
QCD know nothing about QCD, aside from its chiral symmetry. However,
the various models differ in $F_r(T,m)$, so this is the piece which
gives information about the actual degrees of freedom involved in the
QCD phase transition.

\bet\begin{center}\begin{tabular}{|cc|ccc|c|}
  \hline
  Model & Example & $\beta$ & $\delta$ & $\alpha$ & Ref\\ \hline
 O($\infty$) & & 1/2 & 5 & -1 & \cite{largen}\\
 O(4) & chiral QCD & 0.380 & 4.86 & -0.2268 & \cite{engels4} \\
 O(3) & ? & 0.365 & 4.79 & -0.115 & \cite{zinnjustin} \\
 O(2) & liquid He & 0.349 & 4.78 & -0.0172 & \cite{engels2} \\
 O(1) & liquid-gas & 0.325 & 4.8 & 0.11 & \cite{zinnjustin} \\ \hline
 MFT  & & 1/2 & 3 & 0 & \\ \hline
\end{tabular}\end{center}
\caption{Critical exponents of O(N) models in three spatial dimensions.
 The exponent $\alpha$ is obtained from the other reported exponents using
 scaling identities. O(1) should be taken to mean the Ising model. MFT
 stands for mean field theory. There are no known examples of O(3) models,
 since real ferromagnets have relevant terms which break this symmetry.}
\eet{on}

Even if the magnitude of $F_s$ is comparable to $F_r$, since it is
singular, its effect may be enhanced by taking sufficient number of
derivatives. For example, the specific heat, $\cv\propto t^{-\alpha}$,
and hence diverges at $T=T_c$, provided that $\alpha>0$. As one can
see from Table \ref{tb:on}, this is true of the Ising model. However,
for all other $O(N)$ models $\alpha<0$, and, as a result, the singular
contribution to the specific heat exactly vanishes for $T=T_c$.

This seems to contradict our knowledge of the specific heat of liquid He,
which is in the O(2) universality class and has a cusp at the critical
point.  The resolution of this puzzle comes from noticing that the peak
of $\cv$ is only finite, and hence is regular. It is the shape which is
singular. So the specific heat has to arise through a playoff between
the singular and regular parts.  In fact, a very precise microgravity
experiment has been done over the range $|T-T_c|\le2$ nK \cite{lipa},
and the results fitted to the formula
\beq
   \cv = A_r + t^{-\alpha}(B+C t^{-\Delta'}),
\eeq{shape}
where $A_r$ comes only from $F_r$. $\Delta'$ is a possible
correction-to-scaling exponent. $A_r$ is positive and $B$ is
constrained to be negative.  It can be shown that $B$ can be
negative without violating the thermodynamic consistency criterion
that $\cv>0$. A result of the microgravity experiment is that
$\alpha=-0.01285(38)$. Interestingly this is in disagreement with the
careful work of \cite{engels2}.

This mechanism also works for QCD and other O(N) symmetric models. For these
we may write
\beq
   \cv(T,m) = A_r + \frac{t^{-\alpha}}{T_c}\Psi(t,\tau).
\eeq{cvqcd}
$T_c$ and $A_r$ are non-universal, and change from QCD to various
effective theories for it, but the exponent $\alpha$, and the regular
function $\Psi$ (which may be written in terms of the scaling function
$\Phi$, and its derivatives, $\Phi'$ and $\Phi''$, if desired) are
universal. So the shape of the specific heat cusp is universal but its
height and width must be determined in QCD. Furthermore, these two
parameters are good tests of possible effective models, since a bad
model of QCD will not reproduce its non-universal properties.

\bef
\begin{center}
\includegraphics[scale=1.0]{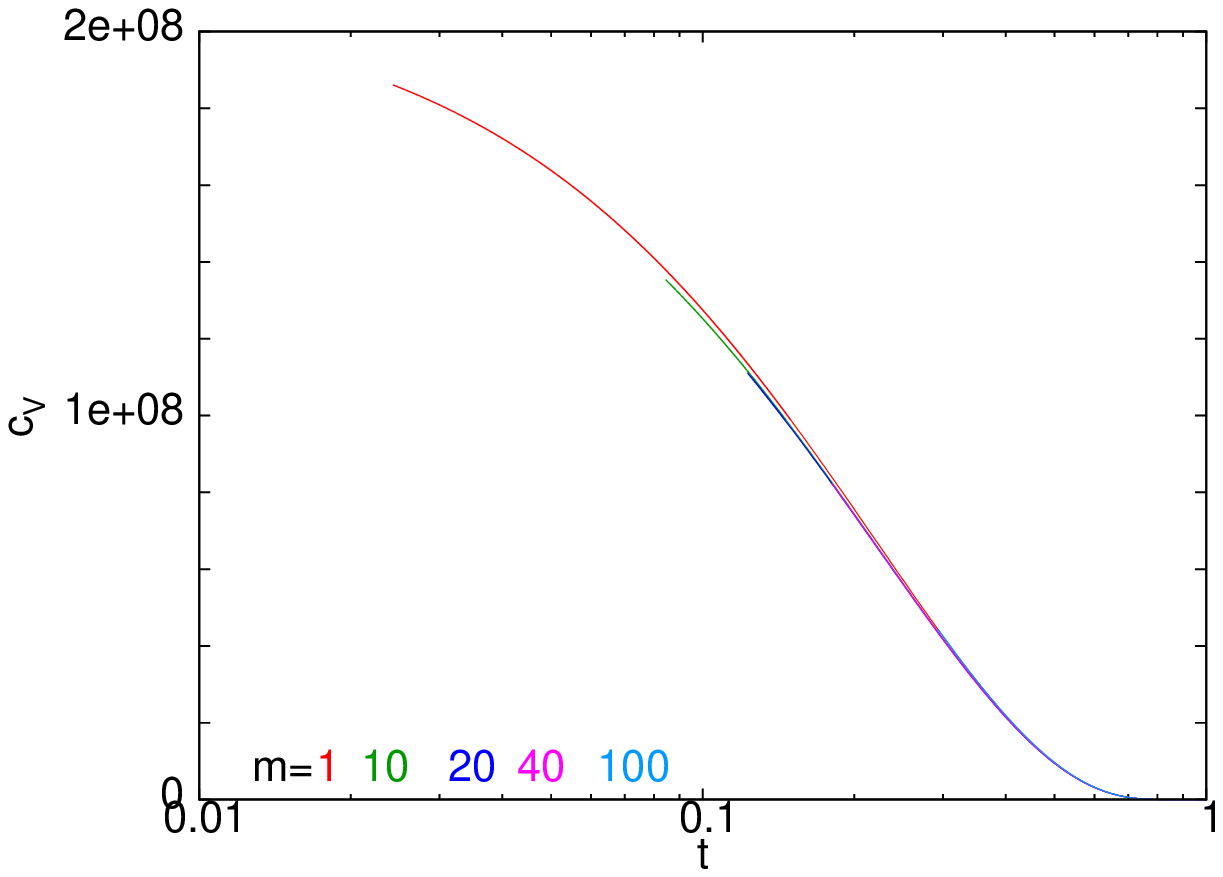}
\end{center}
\caption{Data collapse obtained in an MFT treatment of the NJL model
 when keeping only the data for $\tau>50$. Tiny violations of scaling
 are visible; these can be controlled by increasing the cut on $\tau$.
 The value of this cut depends on the choice of $M_0$ (here it was
 taken to be $T_c$), and the renormalization scheme, when going beyond
 MFT. Also, since this value is not universal, it could be different
 in QCD. Only data for $T<T_c$ is used in this plot to avoid having to
 subtract a large regular part, as discussed in the text.}
\eef{njl}

An interesting statement about the scaling of $\cv$ with mass arises
from this. Suppose we succeeded in measuring (on the lattice) $\cv$ for
QCD with various different light quark masses. By plotting the data as a
function of $t$ and scaling $\cv$ appropriately, can we observe scaling
in the form of data collapsing on to an universal curve? Clearly, there
are no singularities of the free energy if $T$ is varied around $T_c$
at fixed non-vanishing $m$. As a result, taking $t\to0$ and $\tau\to0$
simultaneously will not reveal scaling. Instead, one must take the limit
$m\to0$ first and $T\to T_c$ next, which means that one must take
$\tau\to\infty$ first and then $t\to0$ in order to see data collapse.

One can test this in the NJL model even at tree level, \ie, in the
MFT approximation. The high temperature limit of this model contains
weakly interacting quarks, so the regular contribution to $\cv$ actually
increases fairly rapidly with temperature. As a result, one may miss
the pseudo-critical behaviour in $\cv$ unless the temperature range is
scanned finely to discover a peak sitting over a rising background,
or the free quark contribution is subtracted to make the peak stand
out over a falling background. This difficulty would also occur in QCD
\cite{prasad}, but not in the O(4) Heisenberg magnet.

With this MFT one sees that data collapse is possible when one plots
$\cv$ against $t$ provided that one selects only $\tau>50$. This is
sufficient to ensure that for any finite $m$ one does not approach $t=0$
too closely.  Figure \ref{fg:njl} shows that one may relax the condition
$\tau\to\infty$ provided one is willing to tolerate small enough violation
of scaling. Since experimental data or Monte Carlo computations come with
errors, it should be possible to tune the cutoff on $\tau$ in order to
find the scaling curve within the errors.

We end this section with a remark about the scaling fields. In making
use of effective models to study universal properties of QCD, most works
make the assumption that the scaling fields of the effective theory ($T$
and $m$) are identical to those of QCD. Whether or not this assumption
is correct can be tested, but, to the best of our knowledge, such tests
have not been performed.

\section{Relevance to the phase diagram of chiral QCD}

\bef
\begin{center}
\includegraphics[scale=1.0]{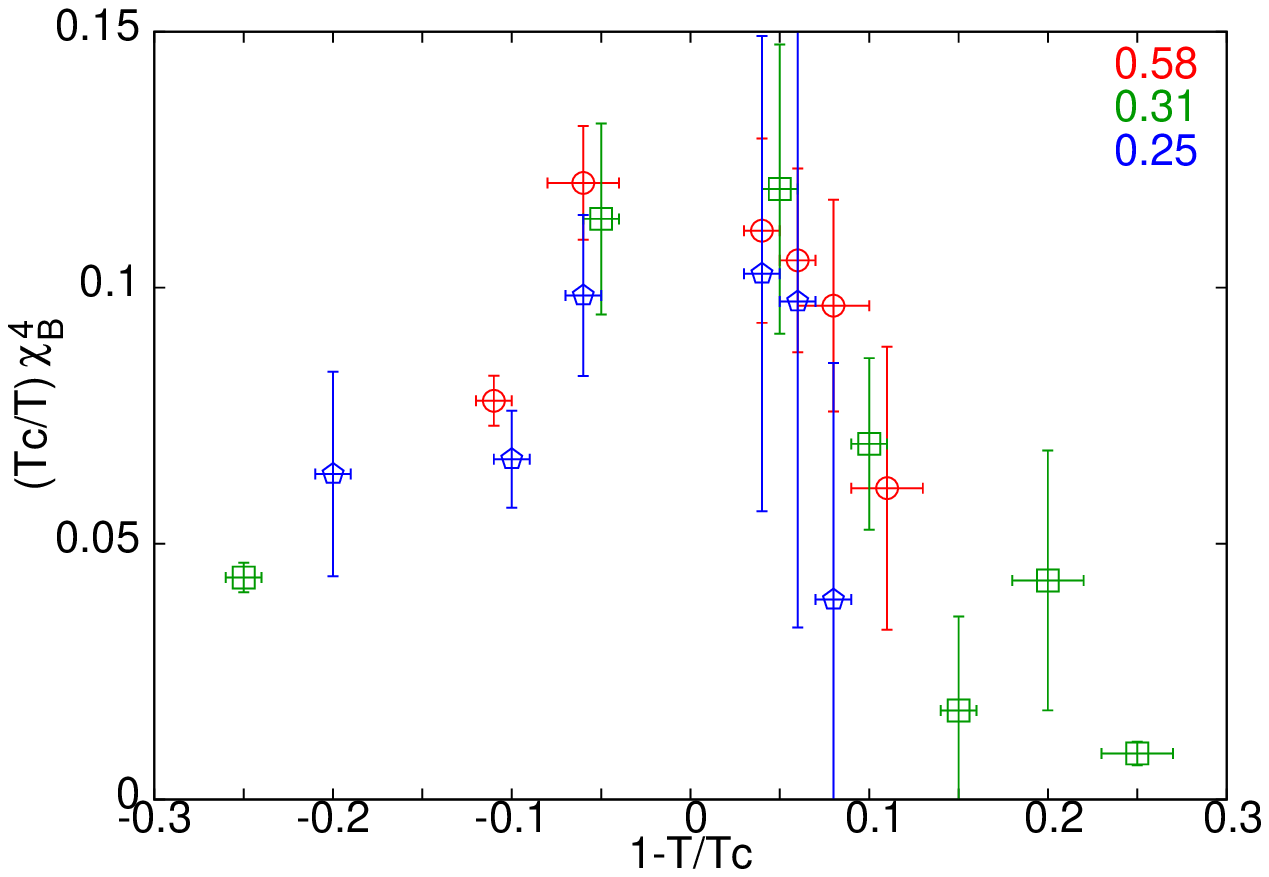}
\end{center}
\caption{Data collapse obtained for $\chi_B^4$ using the results of
 \cite{pushan}, when keeping only the data for $\tau>0.04$ with O(4)
 exponents. The data is plotted as a function of $1-T/T_c$ instead
 of $t$ because the regular parts on the two sides of $T_c$ are have
 not been removed. The colour coding corresponds to the values of
 $m_\pi^2/m_\rho^2$ given in the figure legend.}
\eef{pushan}

The phase diagram of chiral QCD can be extended to finite baryon chemical
potential, $\mu$. Since this scaling field preserves the O(4) symmetry,
the critical point of chiral QCD at $\mu=0$ gets stretched into a line.
The global CP symmetry of QCD implies that $F(T,m=0,\mu)=F(T,m=0,-\mu)$, so
\beq
   T_c(\mu) = T_c + \frac12\kappa\mu^2 + \cdots.
\eeq{kappa}
The curvature has been studied in lattice QCD for about a
decade. Different determinations agree roughly on its value \cite{kappa}.

If one assumes that $\mu$ enters the scaling function of eq.\
(\ref{scaling}) only through the dependence of $T_c$ on $\mu$ as given
in eq.\ (\ref{kappa}), then, as in \cite{krzystoff}, we can write
\beq
   \left.\frac{\partial}{\partial T}\, g(t,\tau)\right|_{\mu=0} = -
      \left.\frac{T_c}{T\kappa}\frac{\partial^2}{\partial\mu^2}\, g(t,\tau)
    \right|_{\mu=0}.
\eeq{ders}
Qualitative evidence for this relation between derivatives was obtained
very long back \cite{patterns}. The relation above implies a connection
between $\cv$ and the fourth order baryon number susceptibility---
\beq
   \chi_B^4(T,m) = \left.
      \frac{\partial^4 P(T,m,\mu)}{\partial\mu^4}\right|_{\mu=0}
      \simeq (\kappa T_c)^2\,\left(\frac T{T_c}\right)^4\,\frac{\cv}{T^3}.
\eeq{divs}
The last expression for $\chi_B^4$ comes from retaining only the most singular
contribution. This suggests a scaling test of $\chi_B^4$ similar to that for
$\cv$.

This test can be performed with the results of \cite{pushan}.  For the
treatment of lattice QCD computations we may replace $m/M_0$ in the
definition of $\tau$ by $m_\pi^2/m_\rho^2$, so that $t$, $\tau$ and
$\chi_B^4$ are all renormalized quantities.  In Figure \ref{fg:pushan} we
plot the full measured $T_c\chi_B^4/T$. Data collapse should be expected
in the region where the contribution from the singular part dominates.

The regular parts at temperatures well above and well below $T_c$ are
expected to be different, since effective theories in these two regions
are the hadron gas model ($T\ll T_c$) and the weak coupling
expansion of QCD ($T\gg T_c$).  We have plotted Figure \ref{fg:pushan}
to show these two branches separately.  It seems that in the region
$t\le0.1$, the differences in the regular parts may be neglected within
the precision of the data. One particular implication is that the
gas model should not work for $\chi_B^4$ within 10\% of $T_c$.

While this gives us a first test (in this sector) of scaling at $T=0$
at surprisingly large quark masses, the current errors are large.
Improvement in errors would allow us to test scaling better, and also
to test the importance of the variation of the regular part of these
quantities with approximately 10\% change in $t$. These requirements
set benchmarks for future measurements of $\chi_B^4$.

It is clear from eq.\ (\ref{scaling}) and the values of $\alpha$ in Table
\ref{tb:on} that derivatives of $\cv$ with respect to $T$ would diverge
in the vicinity of the critical point. As a result, one should be able
to observe scaling of the higher order baryon number susceptibilities.
For example, the sixth order quantity, $\chi_B^6$ would be universal, and
is likely to have a shape similar to that shown in Figure \ref{fg:chi6}.
It would be interesting to test this in future, when improvement in
statistics makes these tests significant.

\bef
\begin{center}
\includegraphics[scale=0.5]{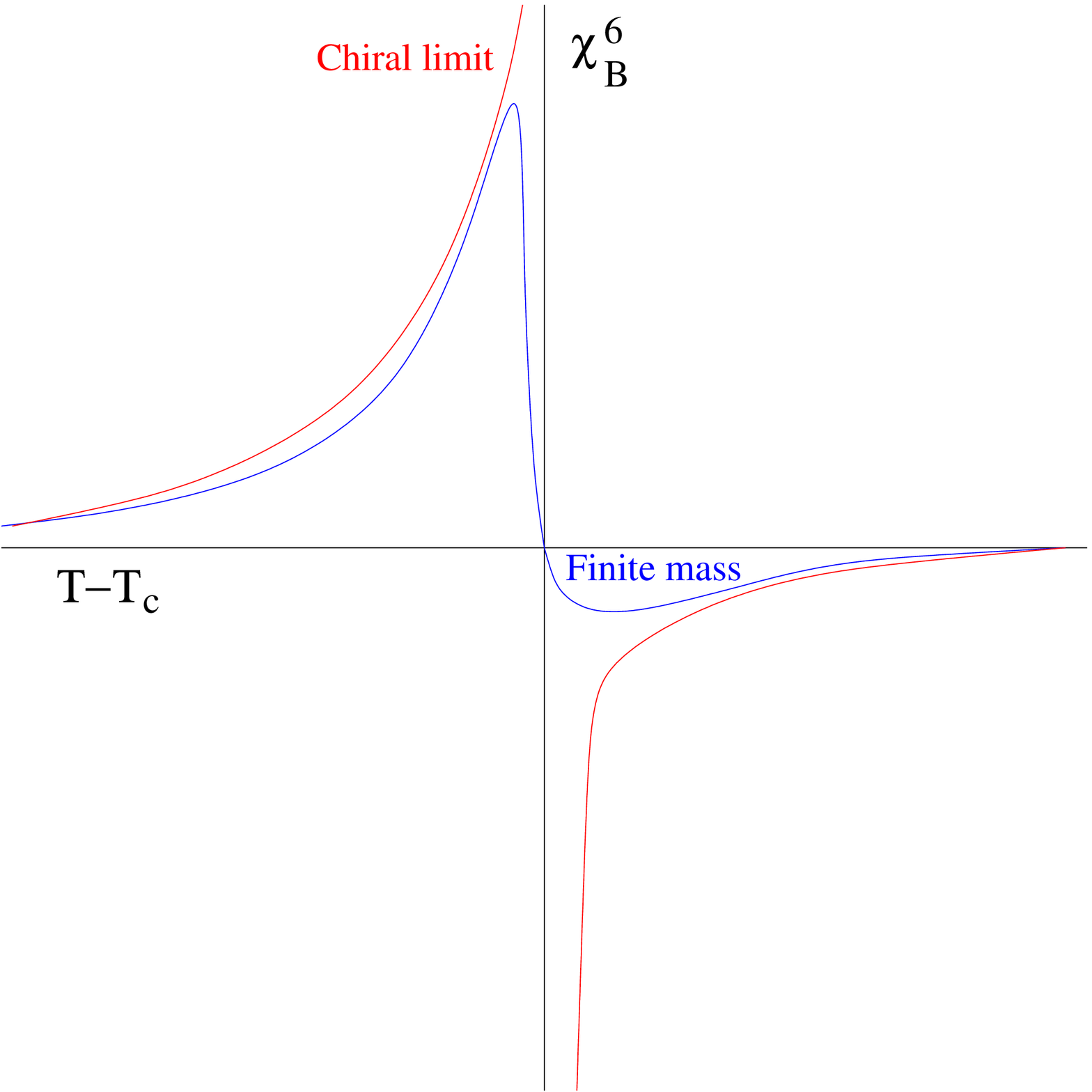}
\end{center}
\caption{In the chiral limit $\chi_B^6$ diverges with critical exponent
 $1+\alpha$ and is proportional to the temperature derivative of $\cv$.
 At finite $m_\pi^2/m_\rho^2$ these divergences would be rounded off
 as shown.}
\eef{chi6}

\section{Some remarks}

It is interesting to recall that before the modern theory of critical
phenomena was developed, the Ehrenfest classification of phase transitions
was in vogue.  This attempted to define orders of phase transitions
according to which derivative of the free energy diverged. In the case of
O(N) models one sees very clearly that such a classification runs into
trouble. On examining the chiral (magnetic) susceptibilities, one comes
to the conclusion that the QCD transition is of second order. However,
on examining $\cv$ one comes to the conclusion that the same transition
is of third order, since $\cv$ does not diverge, but its derivative with
respect to $T$ does. We realize today that the differences are due only
to the value of associated critical index.

We conclude by reiterating the importance of scaling tests such as that
suggested here. They constitute a new domain of tests of the universality
hypothesis in the context of QCD. Not only is this important in its own
right, but also serves to put bounds on the region of applicability of
models such as the hadron gas model.  This model is a mixture of ideal
gases and hence contains no singular part, whereas O(4) universality is
based entirely on the singularity due to pions in the chiral limit. Since
these are mutually exclusive descriptions of the free energy, the success
of one rules out the other.

We thank Deepak Dhar for discussions.

\end{document}